\documentclass[pra,twocolumn,showpacs,aps]{revtex4}
\usepackage{epsfig}
\usepackage{amsmath}
\usepackage{graphicx}
\usepackage{bbm}

\newcommand{\di}{\partial}

\begin{document}

\title{Cooling of mechanical motion with a two level system: the high temperature regime}
\author{P. Rabl}
\affiliation{ITAMP, Harvard-Smithsonian Center for Astrophysics, Cambridge, Massachusetts
02138, USA}
\date{\today}

\begin{abstract}
We analyze cooling of a nano-mechanical resonator coupled to a dissipative solid state two level system focusing on the regime of high initial temperatures. We derive an effective Fokker-Planck equation for the mechanical mode which accounts for saturation 
%of the two level system 
and other non-linear effects and allows us to study the cooling dynamics of the resonator mode for arbitrary occupation numbers. We find a degrading of the cooling rates and eventually a breakdown of cooling at very high initial temperatures and discuss the dependence of these effects on various system parameters. Our results apply to most solid state systems which have been proposed for cooling a mechanical resonator including quantum dots, superconducting qubits and electronic spin qubits.  

\end{abstract}

\maketitle

%85.85.+j 	Micro- and nano-electromechanical systems (MEMS/NEMS) and devices

%Micromechanical devices, 07.10.Cm
%37.10.De 	Atom cooling methods 
%71.55.-i Impurity and defect centers

%\section{Introduction}
Over the past years quantum ground state cooling schemes for micro- and nano-mechanical resonators have attracted considerable attention. 
Work in this direction is motivated by the goal to study quantum effects with macroscopic objects~\cite{ArmourPRL2002,MarshallPRL2003,VitaliPRL2007} as well as potential applications for quantum information processing~\cite{ClelandPRL2004,EisertPRL2004,RablNatPhys2010}  and nano-scale sensing techniques~\cite{MaminAPL2001,MaminNatNano2007,JensenNatNano2008}. Apart from passively cooling high frequency mechanical modes in a cryogenic environment~\cite{ConnellNature2010}  several active cooling schemes are currently explored. So far the most successful techniques are based on opto-mechanical cavity cooling ideas~\cite{OMRevKippenberg,OMRevMarquardt,MarquardtPRL2007,WislonRaePRL2007,GenesPRA2008,YongLiPRB2008,WilsonRaeNJP2008} where the mechanical resonator is damped by interactions with a driven optical~\cite{GroblacherNatPhys2009,ArcizetNAT2006,KlecknerNAT2006,CorbittPRL2007,Schliesser2009, ThompsonNAT2008,Wilson2009} or microwave~\cite{TeufelNJP2008,Rocheleau2009} cavity. Alternatively, mechanical motion can be cooled by coupling the resonator to a dissipative two level system (TLS) and various implementations including quantum dots~\cite{WilsonRaePRL2004,ZippilliPRL2009}, superconducting qubits~\cite{MartinPRB2003,XueNJP2007, ZhangPRL2005,WangNJP2008,JaehneNJP2009,WangPRB2009} and  nitrogen-vacany (NV) defects in diamond~\cite{RablPRB2009}  
have been suggested in this context. Although in practice more challenging such an approach offers the great advantage that once the resonator is cooled to the ground state, the TLS can be used as a non-linear element to prepare and detect non-classical states of the resonator mode.

For cavity cooling schemes the linearized opto-mechanical interactions allow for an exact analytic treatment of the cooling problem~\cite{MarquardtPRL2007,WilsonRaeNJP2008} which in general is not the case for a mechanical resonator coupled to a TLS. Therefore, most theoretical studies of TLS-based cooling schemes~\cite{WilsonRaePRL2004,ZippilliPRL2009,MartinPRB2003,JaehneNJP2009,WangPRB2009,RablPRB2009}  are focused on the regime of weak resonator-TLS interactions which is analogous to the Lamb-Dicke (LD) regime discussed in the context of  laser cooling of trapped atoms and ions~\cite{StenholmRMP1986,CiracPRA1992,LeibfriedRMP2003}. The cooling dynamics is then described by a rate equation for the mean resonator occupation number $n_r(t)$,
\begin{equation}\label{eq:RateEq}
\dot n_r(t)=-\Gamma(n_r(t)-n_f),
\end{equation}
where the damping rate $\Gamma$ and the final occupation number $n_f$ can be calculated from the linear response of the TLS. In this  regime opto-mechanical and TLS-based cooling schemes lead to qualitatively similar predictions for $\Gamma$ and $n_f$ which in essence also agree with the known results from laser cooling of trapped atoms~\cite{StenholmRMP1986,CiracPRA1992,LeibfriedRMP2003}.

Although the weak coupling approximation is usually well justified at low resonator occupation numbers a simple rate equation~\eqref{eq:RateEq} cannot be valid for arbitrary temperatures since the maximum cooling power is naturally bound by the decay rate of the TLS. Recently we have proposed to use optical pumping of the spin states of a NV center in diamond to cool the motion of nano-scale cantilevers~\cite{RablPRB2009}. This setup can in principle be operated at room temperature where for oscillation frequencies of $\omega_r/2\pi\sim1$ MHz the thermal equilibrium occupation number can be as high as $N_{th}\approx 10^6$. Especially in this extreme case one can expect that non-linear effects such as the saturation of the TLS lead to considerable modification of the cooling dynamics. 
%
%
%Therefore, in contrast to cavity cooling schemes, non-linear effects such the  saturation of the TLS can lead to considerable modification of the cooling dynamics, in particular in the initial stage where occupation numbers are close to the thermal equilibrium value $N_{th}$.  
However, also for other physical implementations and lower initial temperatures of $T=0.1-1$ K the resonator energy can still vary over several orders of magnitude during the cooling processes  and the validity of Eq.~\eqref{eq:RateEq} under different experimental conditions must be addressed.

In this work we analyze the cooling dynamics of a mechanical resonator mode coupled to a dissipative TLS, focusing in particular on  the regime of high initial temperatures. %Following related ideas of semi-classical laser cooling theory~\cite{StenholmRMP1986} 
Based on a semi-classical approximation we derive an effective Fokker-Planck equation for the resonator mode which is valid for arbitrary occupation numbers and allows us to study the crossover from the LD to the high temperature regime. Already at moderately high occupation numbers we find a degrading of the cooling rates,  but in this case the final state is still accurately described by the LD theory. However, beyond a certain critical value of $N_{th}$ cooling is highly suppressed and cannot compete with environment-induced re-thermalization processes. In this regime cooling of a mechanical mode with a TLS is no longer possible. We study the crossover  between the different cooling regimes and discuss the dependence of the critical occupation number on the relevant system parameters.

%However, the theoretical framework presented in this paper can also be applied to study related problems including for example the damping of resonator modes coupled to an intrinsic two level fluctuators.      

The paper is structured as follows. In Sec.~\ref{sec:Model} we introduce the basic model for a mechanical resonator coupled to a solid state TLS and present in Sec.~\ref{sec:LambDicke} a brief overview of the cooling results in the LD regime. 
In Sec.~\ref{sec:HighTCoolingRates} we then study the dependence of the cooling rate on the resonator occupation number and derive in Sec.~\ref{sec:FokkerPlanck} an effective Fokker-Planck equation to describe the full cooling dynamics and steady state properties of the system. In Sec.~\ref{sec:MultiMode} we estimate the effects of higher order mechanical modes and finally in Sec.~\ref{sec:Conclusions} we summarize the main results and conclusions of this work.

\section{Model}\label{sec:Model}
We consider a nano-mechanical resonator coupled to a dissipative TLS, for example a superconducting qubit~\cite{MartinPRB2003,ZhangPRL2005,WangNJP2008,XueNJP2007, JaehneNJP2009,WangPRB2009}, a quantum dot~\cite{WilsonRaePRL2004}  or an electronic spin qubit~\cite{RablPRB2009}. We model the resonator as a single harmonic mode with frequency $\omega_r$ and annihilation (creation) operator $a$ ($a^\dag$). The validity of this single mode approximation and the influence of higher order vibrational modes will be addressed in Sec.~\ref{sec:MultiMode}. The lower and upper state $|0\rangle$ and $|1\rangle$ of the TLS are split by a bare transition frequency $\omega_{01}$ and coupled by a classical driving field of frequency $\omega_d=\omega_{01}+\delta$. The motion of the resonator leads to an additional modulation of the TLS splitting and in the frame rotating with $\omega_d$ the total system Hamiltonian is $(\hbar=1)$
\begin{equation}\label{eq:H}
H= -\frac{\delta}{2}\sigma_z + \frac{\Omega}{2}\sigma_x + \omega_r a^\dag a +\frac{\lambda}{2} (a+a^\dag)\sigma_z\,.
\end{equation}
Here $\sigma_k$ are the usual Pauli operators,  $a$ ($a^\dag$) the annihilation (creation) operators for a mechanical mode of frequency $\omega_r$ and   $\Omega$ is the Rabi frequency of the classical driving field. The last term in Eq.~\eqref{eq:H} describes the TLS-resonator coupling where $\lambda$ is the frequency shift per vibrational quanta. Depending on the specific physical implementation the origin of this frequency shift can be due to electrostatic, magnetic or deformation potential interactions, and for a more detailed derivation of Hamiltonian~\eqref{eq:H} we refer the reader to the original proposal cited above.

\subsection{Master equation}
The TLS and the resonator mode interact with the environment and we model the resulting dissipative dynamics by a master equation
\begin{equation}\label{eq:ME}
\dot \rho=-i[H,\rho] + \mathcal{L}_\Gamma  (\rho) + \mathcal{L}_\gamma(\rho),
\end{equation} 
for the system density operator $\rho$. The Liouville operator $\mathcal{L}_{\Gamma}$  accounts for relaxation and dephasing processes of the TLS %with rates $\Gamma_\perp$ and $\Gamma_\parallel$ respectively
and % by including the effects of a finite equilibrium population of state $|1\rangle$ it 
 is in general given by   
\begin{equation}
\begin{split}
 \mathcal{L}_\Gamma  (\rho) =& \frac{\Gamma_\perp}{2} (N_{\rm TLS}+1) \left(2\sigma_- \rho \sigma_+ -\sigma_+\sigma_- \rho -\rho \sigma_+\sigma_-\right)\\
 +&\frac{\Gamma_\perp}{2} N_{ \rm TLS} \left(2\sigma_+ \rho \sigma_- -\sigma_-\sigma_+ \rho -\rho \sigma_-\sigma_+\right)\\
 +&\frac{\Gamma_\parallel}{2}\left(\sigma_z\rho\sigma_z-\rho\right).
 \end{split}
 \end{equation}
Here the first two lines describe relaxation of the TLS  towards the equilibrium value $\langle \sigma_z\rangle(t\rightarrow\infty)=-1/(2N_{\rm TLS}+1)$ with a total rate  $T_1^{-1}=\Gamma_\perp(2N_{\rm TLS}+1)$. Note that the effective occupation number may differ from the thermal equilibrium value $N_{\rm TLS}=(e^{\hbar \omega_{01}/k_BT}-1)^{-1}$ in the presence of experimental imperfections. The third line accounts for additional dephasing processes and the total dephasing rate of the TLS is  $T_2^{-1}=T_1^{-1}/2+\Gamma_\parallel$. In case the TLS is represented by the spin states of a NV center the effective decay rate $\Gamma_\perp$ is adjustable by the optical spin pumping rate and $\Gamma_\parallel, N_{\rm TLS}>0$ are related to non-ideal optical pumping processes~\cite{RablPRB2009}.   

The third term in Eq.~\eqref{eq:ME} describes mechanical dissipation due interactions with a thermal phonon reservoir of the support or other intrinsic damping mechanisms. We model mechanical dissipation by
\begin{equation}
\begin{split}
 \mathcal{L}_\gamma  (\rho) = &\frac{\gamma}{2}(N_{th}+1) \left(2a \rho a^\dag -a^\dag a \rho -\rho a^\dag a\right)\\ &\frac{\gamma}{2}N_{th} \left(2a^\dag \rho a -a a^\dag \rho -\rho a a^\dag \right),
\end{split}
 \end{equation}
where $\gamma=\omega_r/Q$ is the energy damping rate for a mechanical quality factor $Q$ and $N_{th}=(e^{\hbar \omega_r/k_BT}-1)^{-1}$ is the equilibrium phonon occupation number. Throughout this work we focus on the high $Q$ and high temperature regime where $\gamma N_{th}\simeq k_BT/\hbar Q$ is independent of the resonator frequency.

\subsection{Displaced oscillator basis $\&$ the Lamb-Dicke parameter}
%Our goal in the following is to study effective cooling dynamics of the resonator state for the model defined by 
Our goal in the following is to study the effective cooling dynamics of the resonator mode which results from interactions with the dissipative TSL. Before we proceed let us briefly remark on the relation between the TLS-resonator coupling defined by Eq.~\eqref{eq:H}
%The TLS-resonator coupling defined Eq.~\eqref{eq:H} appears frequently in the description of mechanical resonators coupled to solid state TLSs and before we proceed let us briefly remark on its relation
%The model defined in Eq.~\eqref{eq:H} appears frequently in the description of mechanical resonators coupled to solid state TLSs also  %in the context of cooling. Therefore, before we proceed let us briefly remark on its relation
%
%to 
and the basic model for atom-light interactions,  
\begin{equation}\label{eq:HAL}
H_{A-L}=\frac{\Omega_L}{2} \left(e^{i\eta_{LD} (a^\dag+a)}\sigma_++ e^{-i\eta_{LD} (a^\dag+a)}\sigma_-\right),
\end{equation}
discussed in the context of laser cooling of trapped atoms and ions~\cite{StenholmRMP1986,CiracPRA1992,LeibfriedRMP2003}. Eq.~\eqref{eq:HAL} describes the interaction of a trapped two level atom with a laser of Rabi frequency $\Omega_L$. The displacement operators account for the photon recoil whenever the atomic state is flipped. Here
$\eta_{LD}=k_La_0$ is the Lamb-Dicke parameter for a  laser wavevector $k_L$ and $a_0$ the size of the groundstate wavefunction of the trapped atom. It can be re-expressed as $\eta_{LD}=\sqrt{\omega_{\rm recoil}/\omega_t}$ where $\omega_{\rm recoil}=\hbar k_L^2/2m$ is the recoil  and $\omega_t$ the trapping frequency. For a tight confinement and low occupation numbers, $\eta_{LD} \sqrt{\langle a^\dag a\rangle} \ll 1$, Eq.~\eqref{eq:HAL} can be expanded to lowest order in $\eta_{LD}$ which, for example, leads to a simplified description of sideband cooling schemes for trapped ions~\cite{CiracPRA1992,LeibfriedRMP2003}. For weak confinement or high temperatures this expansion is no longer possible. In this case $\varepsilon=\omega_{\rm recoil}/\Gamma_\perp=\eta_{LD}^2 \omega_t/\Gamma_\perp$ has been identified as the relevant expansion parameter which for $\varepsilon\ll 1$  still allows for a semiclassical treatment of laser cooling problems~\cite{StenholmRMP1986}.  

As pointed out, e.g., in Ref.~\cite{WilsonRaePRL2004} the connection between Eq.~\eqref{eq:H} and Eq.~\eqref{eq:HAL} is most apparent in the displaced oscillator basis which is defined by the unitary (polaron) transformation $\tilde H= UHU^\dag$ where $U=e^{iS}$ and $S=-i\lambda/(2\omega_r)(a^\dag-a)\sigma_z$. In this basis  we obtain 
\begin{equation}\label{eq:Hpolaron}
\tilde H = -\frac{\delta}{2}\sigma_z + \frac{\Omega}{2}\left(e^{\frac{\lambda}{\omega_r} (a^\dag-a)}\sigma_+ +e^{-\frac{\lambda}{\omega_r}(a^\dag-a)} \sigma_-\right) +\omega_r a^\dag a\,,
\end{equation}
%\begin{figure}
%\begin{center}
%\includegraphics[width=0.25\textwidth]{Displaced}
%\caption{Displaced oscillator basis. }\label{Displaced}
%\end{center} 
%\end{figure}
and see that in analogy to the momentum kick in Eq.~\eqref{eq:HAL} a flip of the TLS state is now  accompanied by a displacement of the resonator state in position space. By comparing Eq.~\eqref{eq:Hpolaron} and Eq.~\eqref{eq:HAL} we find that in the present system $\eta=\lambda/\omega_r$ is the equivalent of the LD parameter while the quantity $\lambda^2/\omega_r$ can be identified with the recoil frequency.  Therefore, we can already at this stage argue that for the experimentally most relevant regime 
 $\lambda \ll \omega_r,\Gamma_\perp$ an approximate treatment of the system dynamics in close analogy to semiclassical laser cooling problems should also be applicable for the present system.

\section{The Lamb-Dicke regime}\label{sec:LambDicke}
%Our goal in the following is to study the effective cooling dynamics of the resonator mode which results from interactions with the dissipative TSL.
As a starting point we first consider in this section the weak coupling \emph{and} low temperature regime where $\lambda \sqrt{N_{th}+1/2} \ll \Gamma_{\perp},\omega_r$ is satisfied. % In this regime the interaction Hamiltonian $H_\lambda=\frac{\lambda}{2}(a+a^\dag)\sigma_z$ 
As discussed above, this limit is equivalent to the LD regime in laser cooling and has also been analyzed in several previous works on cooling schemes for mechanical resonators~\cite{WilsonRaePRL2004,ZippilliPRL2009,MartinPRB2003,JaehneNJP2009,WangPRB2009,RablPRB2009}. Here we only present a brief summary of the main results in this limit which we will be the basis for further discussions in Sec.~\ref{sec:HighTCoolingRates} and Sec.~\ref{sec:FokkerPlanck}.

\subsection{Rate equation in the LD regime}
In the LD regime the effect of the resonator-TLS interaction can be treated as a small perturbation and it is convenient to decompose the  master equation~\eqref{eq:ME} into three terms
\begin{equation}\label{eq:MELD}
\dot \rho=\mathcal{L}_{\rm TLS}(\rho)+\mathcal{L}_r(\rho)+\mathcal{L}_\lambda(\rho), 
\end{equation}
such that $\mathcal{L}_{\rm TLS}$ and $\mathcal{L}_{r}$
%\begin{eqnarray}
%\mathcal{L}_{TLS}(\rho)&=&-\frac{i}{2}[-\delta \sigma_z+\Omega \sigma_x,\rho]+\mathcal{L}_\Gamma(\rho),\\
%\mathcal{L}_{r}(\rho)&=&-i[\omega_r a^\dag a,\rho]+\mathcal{L}_\Gamma(\rho),
%\end{eqnarray}
describe the unperturbed evolution of the TLS and the resonator respectively and 
\begin{equation}\label{eq:Lint}
\mathcal{L}_\lambda(\rho) =-i\frac{\lambda}{2} [(a   +a^\dag ) \Delta \sigma_z,\rho],
\end{equation}
accounts for the coupling. In Eq.~\eqref{eq:Lint} we have set $\Delta \sigma_z=\sigma_z-\langle \sigma_z\rangle_0$ and omitted a constant force proportional to the steady state value $\langle \sigma_z\rangle_0$.  This force does not contribute to cooling and can be reabsorbed in a shift of the resonator equilibrium position.
Under the action of $\mathcal{L}_{\rm TLS}$ the TLS relaxes to its steady state $\rho_{\rm TLS}^0$ on a time scale $T_1$ which is fast 
compared to the timescale associated with $\mathcal{L}_\lambda$. Therefore, the total system density operator can be to a good approximation written as $\rho(t)\simeq\rho_{\rm TLS}^0\otimes \rho_{r}(t)$ and the effects of $\mathcal{L}_\lambda$ on the dynamics of the resonator state $\rho_r(t)$ can be evaluated in second order perturbation theory~\cite{CiracPRA1992}. For the present system this calculation is detailed in Ref.~\cite{JaehneNJP2009}  and as a result we obtain an effective master equation which in the frame rotating with $\omega_r$ can be written as
\begin{equation}\label{eq:CoolingEquation}
\begin{split}
\dot \rho_r=  \mathcal{L}_\gamma(\rho_r)+&\frac{\Gamma_c }{2}(N_0+1) \left(2a \rho_r a^\dag -a^\dag a \rho_r -\rho_r a^\dag a\right)\\+
&\frac{\Gamma_c}{2} N_0 \left(2a^\dag \rho_r a -a a^\dag \rho_r -\rho_r a a^\dag \right).
\end{split}
\end{equation}
Here we have introduced a cooling rate  $\Gamma_{c}=S(\omega_r)-S(-\omega_r)$  and the minimal occupation number $N_0=S(-\omega_r)/\Gamma_c$ which are determined by the equilibrium fluctuation spectrum 
\begin{equation}\label{eq:Spectrum}
S(\omega)=\frac{\lambda^2}{2} {\rm Re} \int_0^\infty d\tau\, ( \langle \sigma_z(\tau)\sigma_z(0)\rangle_0- \langle \sigma_z\rangle_0^2)\,e^{i\omega \tau}\,.
\end{equation}
From the effective master equation~\eqref{eq:CoolingEquation} we can now  derive the rate equation~\eqref{eq:RateEq} for the resonator occupation number $n_r(t)={\rm Tr}\{ a^\dag a \rho_r(t)\}$
%\begin{equation}
% \dot n_r(t) =-\Gamma (  n_r(t) - n_{0,LD}),
%\end{equation}
with a total damping rate $\Gamma=\Gamma_{c}+\gamma$ and $n_f=n_{LD}$. Here
\begin{equation}
n_{LD} \simeq \frac{\gamma N_{th}}{\Gamma_c}+N_{0},
\end{equation}
is the steady state occupation number in the LD limit.

\begin{figure}
\begin{center}
\includegraphics[width=0.4\textwidth]{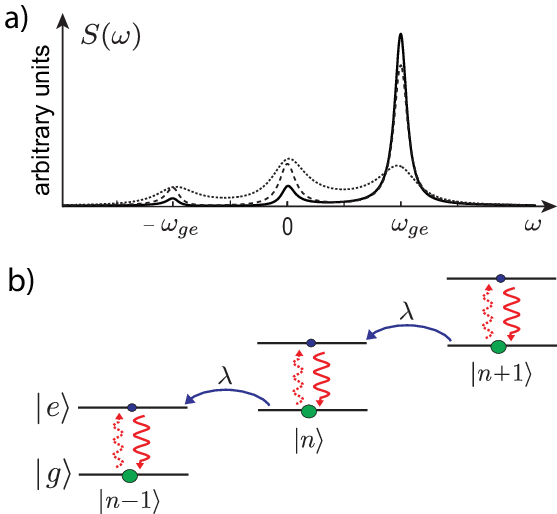}
\caption{a) Fluctuation spectrum $S(\omega)$ where $\omega_{ge}=\sqrt{\delta^2+\Omega^2}$ is the frequency splitting between the two dressed states $|g\rangle$ and $|e\rangle$. The solid line shows the result for $\Omega=0.75 \omega_r$, $\Gamma_\perp=0.1\omega_r$ and the ideal case $\Gamma_\parallel=N_{\rm TLS}=0$. The other two curves indicate the effects of a finite thermal population $N_{\rm TLS}=0.15$ (dashed line) and excess dephasing $\Gamma_\parallel=2\Gamma_\perp$ (dotted line).  b) Level scheme of the coupled resonator-TLS under resonance conditions $\omega_r=\omega_{ge}$, where  $|n\rangle$ denote the vibrational eigenstates. }\label{Figure1}
\end{center} 
\end{figure} 

In the LD regime the cooling dynamics of the resonator mode is fully determined by the fluctuation spectrum $S(\pm \omega_r)$ which characterizes the ability of the \emph{unperturbed} TLS to absorb or emit energy at the mechanical frequency $\omega_r$.  Fig.~\ref{Figure1} shows the typical behavior of $S(\omega)$ together with the energy level diagram which qualitatively explains the relevant cooling and heating processes.
The resonance at $\omega=\omega_{ge}=\sqrt{\delta^2+\Omega^2}$ corresponds to a transition between $|g\rangle\rightarrow|e\rangle$ which are the two dressed eigenstates of the driven TLS, 
\begin{eqnarray*}
|g\rangle &=& \cos(\theta/2) |0\rangle - \sin(\theta/2) |1\rangle,\\
 |e\rangle &=& \sin(\theta/2) |0\rangle + \cos(\theta/2) |1\rangle,
\end{eqnarray*}
where $\theta= \arctan(\Omega/|\delta|)$. This process leads to cooling and in the sideband resolved regime $\omega_{ge}T_2\gg1 $ it is most effective when $\omega_r\approx \omega_{eg}$.  For a finite steady state population in the state $|e\rangle$ the reverse process $\sim S(-\omega_{eg})$ leads to heating. This occurs for $N_{\rm TLS},\Gamma_\parallel>0$ or under very strong driving conditions $\Omega\gg \delta$.  In the Doppler regime $T_2\omega_{ge}<1$ the cooling and heating resonances start to overlap and also pure diffusion processes associated from the resonance at $\omega=0$ become important. In this regime cooling is less efficient, but as we see below also more robust.    

The spectrum $S(\omega)$ can be evaluated using the quantum regression theorem~\cite{
QRT} which allows us to calculate the cooling rate $\Gamma_c$ and the minimal occupation number $N_0$ for arbitrary system parameters.  The general expressions are given in App.~\ref{app:A} and in the following we summarize the main results.

\subsection{Final occupation numbers}
In the limit of an isolated resonator mode $\gamma\rightarrow 0$ the final occupation number is limited by $n_{LD}\simeq N_0$. For a pure decay process, i.e. $\Gamma_\parallel,N_{\rm TLS}\rightarrow 0$, we find that $N_0$ is independent of the Rabi-frequency $\Omega$ and by choosing an optimal detuning $\delta=-\sqrt{\omega_r^2+\Gamma_\perp^2/4}$ we obtain 
\begin{equation}\label{eq:nideal}
 N_0=\frac{1}{2}\left(\sqrt{1+\frac{\Gamma_\perp^2}{4\omega_r^2}}-1\right).
\end{equation}
In the two limiting cases $\Gamma_\perp \ll\omega_r$ and $\Gamma_\perp\gg\omega_r$ Eq.~\eqref{eq:nideal} reproduces the well known results for sideband and Doppler cooling respectively. In particular, this means that $\Gamma_\perp<\omega_r$ is a minimal requirement to achieve ground state cooling. In a solid state environment excess dephasing processes $\Gamma_\parallel>0$ are often non-negligible and cause an additional broadening of the TSL transition. While in this case the detailed behavior of $N_0$ is more complicated, we find that in essence ground state cooling requirements are now determined by the condition $T_2 \omega_r >1$~\cite{WangPRB2009} as expected from the qualitative discussion given above.
Finally, finite temperature effects or other imperfections can cause a non-vanishing occupation number $N_{\rm TLS}$ which introduces  an additional limit $N_0>N_{\rm TLS}$ on the minimal occupation number. More over, we find  that for weak driving and sideband resolved conditions
\begin{equation}
N_0\approx N_{\rm TLS} + \frac{\Gamma_\perp^2}{16\omega_r^2}+  N_{\rm TLS}\frac{\Gamma_\perp^2}{\Omega^2} + \mathcal{O}(N_{\rm TLS}^2),
\end{equation}
which leads to a divergent occupation number for $\Omega\rightarrow0$. This can be understood from the fact that thermally activated jumps between the states $|0\rangle$ and $|1\rangle$ cause a fluctuating force, which for $\Omega\rightarrow0$ is not compensated by a corresponding damping mechanism. This effect does usually not appear in atomic laser cooling, but has been discussed, e.g., for microwave cooling schemes for polar molecules~\cite{WallquistNJP2008}.

In summary we see that ground state cooling can in principle be achieved under sideband resolved conditions  $\Gamma_\perp, \Gamma_\parallel< \omega_r$, low occupation numbers $N_{\rm TLS}<1$ and a minimal driving strength $\Omega > N_{\rm TLS} \Gamma_\perp$. For most implementations discussed in  the literature these conditions can be satisfied. Therefore, we expect that in many initial  experiments the final occupation number will not be limited by $N_0$, but rather by the competition
between cooling and thermalization processes. In this case $n_{LD} \simeq \gamma N_{th}/\Gamma_c$ and therefore in the rest of the paper we will mainly focus on the cooling rate $\Gamma_c$.

\subsection{Cooling rates}
\begin{figure}
\begin{center}
\includegraphics[width=0.48\textwidth]{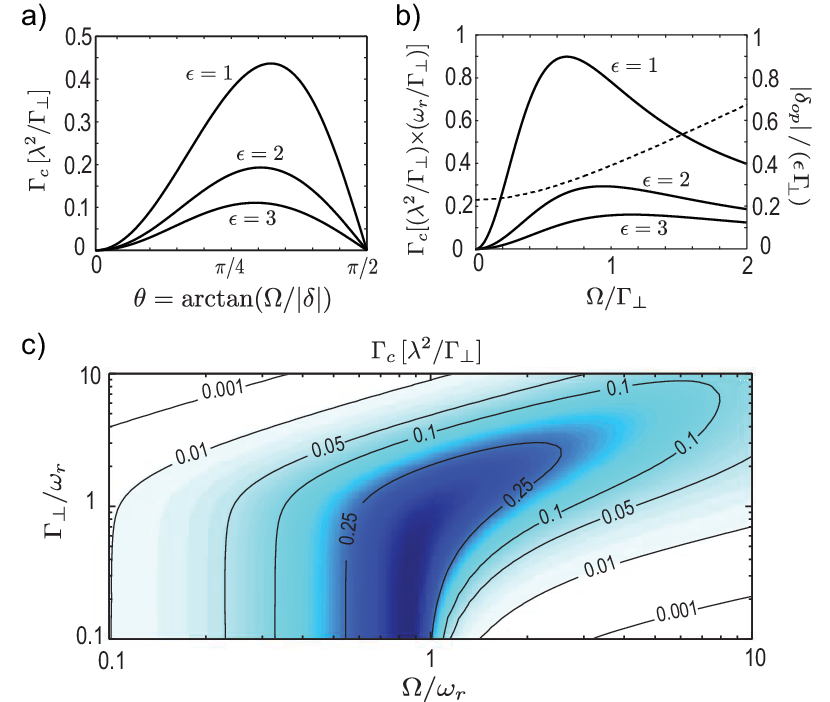}
\caption{Cooling rates in the LD regime. a) Cooling rate $\Gamma_c$ in the sideband limit $\omega_rT_2\gg1$ for different $\epsilon=2T_1/T_2$ and $\omega_r=\sqrt{\delta^2+\Omega^2}$.  b) Cooling rate $\Gamma_c$ in the Doppler limit $\omega_rT_2\ll1$. For each value of  $\Omega$ and $\epsilon$ the detuning is numerically optimized as indicated by the dashed line for $\epsilon=1$.  c) Cooling rate $\Gamma_c$ as a function of $\Gamma_\perp$ and $\Omega$ for the ideal case $\epsilon=1$. For each parameter set the detuning $\delta$ is optimized to maximize $\Gamma_c$.  }\label{fig:CoolingRatesLD}
\end{center} 
\end{figure}

Let us now study the cooling rate $\Gamma_c$ which in general is maximized for different parameters than $N_0$ is minimized. For the following discussion we can assume $N_{\rm TLS}\approx 0$~\cite{NoteNTLS} and we introduce a parameter $\epsilon=2T_1/T_2\geq 1$ which characterizes the excess dephasing compared to the bare decay rate. We have already argued above that in the sideband resolved regime $ T_2\omega_r >1 $ cooling is optimized when $\omega_r$ matches the eigenfrequency $\omega_{ge}=\sqrt{\delta^2+\Omega^2}$
 of the dressed TLS.  Therefore,  we can set $
\delta=-\omega_r\cos(\theta)$ and $\Omega=\omega_r\sin(\theta)$ for $\theta\in [0,\pi/2]$. 
We then obtain 
\begin{equation}
\Gamma_{c}\simeq\frac{\lambda^2}{\Gamma_\perp}\frac{2 \sin(\theta)\sin(2\theta)}{(2\cos^2(\theta)\!+\!\epsilon \sin^2(\theta))(2\sin^2(\theta)\!+\!\epsilon\!+\!\epsilon \cos^2(\theta) )},
\end{equation}
and similar expressions have been derived in Ref.~\cite{JaehneNJP2009} for the case $\epsilon=1$ and in Ref.~\cite{RablPRB2009} for a three level system. In Fig.~\ref{fig:CoolingRatesLD} a) we plot $\Gamma_c$ for different parameters. The maximal cooling rate of $\Gamma_{c}\approx 0.44 \times \lambda^2/\Gamma_\perp$ is achieved for $\epsilon=1$ and $\theta\approx \pi/3$, which corresponds to a driving strength of $\Omega\approx 0.85 \omega_r$~\cite{JaehneNJP2009}.

In the Doppler regime $T_2 \omega_r\ll 1$ the transition frequency of the TLS can no longer be resolved and in this case the cooling rate is optimized for a detuning $\delta\approx -1/(4T_2)$. In Fig.~\ref{fig:CoolingRatesLD} b) we plot $\Gamma_c$ in the Doppler regime as a function of $\Omega$. We find that compared to the sideband limit the maximal cooling rate is reduced by another factor of $\omega_r T_2$ and the maximal cooling rate is achieved for a driving strength $\Omega\approx T_2^{-1}$. Finally, Fig.~\ref{fig:CoolingRatesLD} c) shows the cooling rate as a function of $\Gamma_\perp$ and $\Omega$ for the ideal case $\epsilon=1$. From this plot we conclude that cooling rates close to the maximum value of $\Gamma_c \sim \mathcal{O}(\lambda^2/\Gamma_\perp)$ can be achieved for decay rates up to $\Gamma_\perp\sim \omega_r$, but only under strong driving conditions $\Omega\sim {\rm max}\{\omega_r,\Gamma_\perp\}$.

\section{Cooling rates in the high temperature regime}\label{sec:HighTCoolingRates}
Our goal is now to extend the previous cooling results to the regime of high temperatures where  the condition $\lambda\sqrt{N_{th}}\ll \Gamma_\perp$ is violated.  
%
%
%In this case a simple perturbative treatment of the interaction Hamiltonian $H_\lambda=\frac{\lambda}{2}(a^\dag+a)\sigma_z$ is no longer possible. To proceed it is instructive to look at the effect of $H_\lambda$ on the dynamics of the TLS and the resonator mode respectively, 
%\begin{eqnarray}
%\dot \sigma_- &=& i [H_\lambda,\sigma_-] \sim \frac{\lambda}{2} (a+a^\dag),\\
%\dot a &=& i [H_\lambda,a] \sim \frac{\lambda}{2} \sigma_z.
%\end{eqnarray}
%We see that the action of the resonator motion on the TLS is $\sim \mathcal{O}(\lambda \sqrt{N_{th}})$ and can substantially modify the dynamics of the TLS. However, also at high temperatures the effect of the TLS on the resonator dynamics is $\sim\mathcal{O}(\lambda)$ and can still be treated as a small perturbation. In the following we use this basic insight to derive a semi-classical model for the cooling dynamics which is valid in the regime $\lambda^2/(\omega_r\Gamma_\perp) \ll 1$, but otherwise does not make any assumptions about the parameter $\lambda\sqrt{N_{th}}$. 
%
%As a first step we will in this section study the dependence of the cooling rate on the resonator occupation number, while a more rigorous discussion of the full cooling dynamics is discussed in Sec.~\ref{sec:FokkerPlanck}.
%
As a first step we will in this section use a simple semi-classical approximation to study the dependence of the cooling rate on the mean resonator occupation number. A  more rigorous discussion of the full cooling dynamics and the steady state of the system is then presented in Sec.~\ref{sec:FokkerPlanck}.

\subsection{Semi-classical cooling rates}
Let us change into a frame rotating with $\omega_r$ and for the moment assume that the resonator is prepared in a coherent state $\rho_r=|\alpha\rangle\langle \alpha|$ with $|\alpha|^2\sim N_{th}\gg 1$. In the limit $\gamma\rightarrow 0$ the change in the occupation number is then given by 
\begin{equation}\label{eq:dtadaga}
\di_t\langle a^\dag a\rangle = -i\frac{\lambda}{2}\left(\alpha^* e^{i\omega_r t}-\alpha e^{-i\omega_r t}\right) \langle \sigma_z(t)\rangle.
\end{equation}
We introduce the Bloch vector $\vec S=(\sigma_-,\sigma_+,\sigma_z)^T$ and evaluate its dynamics using a semi-classical approximation, e.g. $\langle a \sigma_k\rangle\approx \langle a\rangle \langle \sigma_k\rangle$. As  a result we obtain modified Bloch equation which in a matrix notation can be written as
\begin{equation}\label{eq:DrivenBloch}
\langle \dot{\vec S}\rangle= {\bf A}  \langle \vec S\rangle  - i \lambda \left( e^{-i\omega_r t}\alpha  + e^{i\omega_r t}\alpha^* \right){\bf A}_z\langle \vec S\rangle - \Gamma_\perp \vec V_z,
\end{equation}
where  $ \vec V_z=(0,0,1)^T$ and $({\bf A}_z)_{11}=- ({\bf A}_z)_{22}=1$ and $({\bf A}_z)_{ij}=0$ otherwise. The first term in Eq.~\eqref{eq:DrivenBloch}  is the free TLS evolution where 
\begin{equation}\label{eq:AMatrix}
  {\bf A}= \left(\begin{array}{ccc}
i\delta-\frac{1}{T_2} & 0 & i\Omega/2 \\
0& -i\delta-\frac{1}{T_2} & - i\Omega/2 \\
i\Omega & -i\Omega & - \frac{1}{T_1} 
\end{array}\right),
\end{equation}
and has eigenvalues of $\mathcal{O}(T_1^{-1})$. The second term in Eq.~\eqref{eq:DrivenBloch} describes an additional driving field  $\sim \lambda \alpha$ caused by the resonator motion. To proceed we use the continued fraction method developed by Stenholm and Lamb~\cite{StenholmPR1969,MinoginOpCom1979}
%by Stenholm and Lamb \cite{StenholmPR1969,MinoginOpCom1979}
%use the Fourier series and continued fraction method developed
%by Stenholm and Lamb \cite{StenholmPR1969,MinoginOpCom1979} and 
and write the Bloch vector in a Floquet representation
\begin{equation}\label{eq:Floquet}
\langle \vec S\rangle (t)= \sum_{n=-\infty}^{\infty} \vec S^n(t) e^{-i n \omega_r t}.
\end{equation}
Here $\vec S^n(t)$ are slowly varying coefficients which relax to their steady state values $\vec S^n_0$ 
on a timescale $T_1$. For $\lambda T_1 \ll 1$ the resonator amplitude does not considerably change on this timescale and in Eq.~\eqref{eq:DrivenBloch} we can treat $\alpha$ as a constant parameter. Then, by inserting Eq.~\eqref{eq:Floquet} into Eq.~\eqref{eq:DrivenBloch} we can express the steady state values of $S_{z,0}^{\pm1}(\alpha)$ in terms of a matrix continued fraction
%~\cite{StenholmPR1969,MinoginOpCom1979}
\begin{equation}\label{eq:ContFrac}
-i \alpha \lambda {\bf A}_z  \vec S^{n-1}_0 +  ({\bf A}+ i\omega_r n\mathbbm{1}) \vec S^n_0-  i \alpha^*  \lambda{\bf A}_z \vec S^{n+1}_0
 =  \delta_{n,0}\Gamma_\perp \vec V_z,
\end{equation}
which can be efficiently evaluated numerically. After inserting the steady state values back into Eq.~\eqref{eq:dtadaga} and taking the average over a few oscillation periods we obtain 
\begin{equation}\label{eq:SlowCooling}
\di_t\langle a^\dag a\rangle \simeq -i \frac{\lambda}{2}\left(\alpha^* S_{z,0}^{+1}(\alpha)-\alpha S_{z,0}^{-1}(\alpha)\right) =-\Gamma_c(\alpha) |\alpha|^2,
\end{equation}
where we have defined the amplitude dependent cooling rate  $\Gamma_c(\alpha):= i\lambda ( S_{z,0}^{1}/\alpha-S_{z,0}^{-1}/\alpha^*)$ which depends on $|\alpha|$ only. 
%By inserting Eq.~\eqref{eq:Floquet} into Eq.~\eqref{eq:DrivenBloch} we find that the steady state values of $S_{z,0}^{\pm1}(\alpha)$ can be expressed in terms of a matrix continued fraction~\cite{StenholmPR1969,MinoginOpCom1979}
%\begin{equation}\label{eq:ContFrac}
%-i \alpha \lambda {\bf A}_z  \vec S^{n-1}_0 +  ({\bf A}+ i\omega_r n\mathbbm{1}) \vec S^n_0-  i \alpha^*  \lambda{\bf A}_z \vec S^{n+1}_0
% =  \delta_{n,0}\Gamma_\perp \vec V_z,
%\end{equation}
%which can be efficiently solved numerically. 
Note that in deriving Eq.~\eqref{eq:SlowCooling} we have assumed that $\lambda \ll \Gamma_\perp$, but also that the mean occupation number does not considerably change over one oscillation period. The self-consistency of this approximation requires $\lambda^2/(\Gamma_\perp \omega_r)\ll 1$.

\subsection{Results $\&$ discussion}
\begin{figure}
\begin{center}
\includegraphics[width=0.48\textwidth]{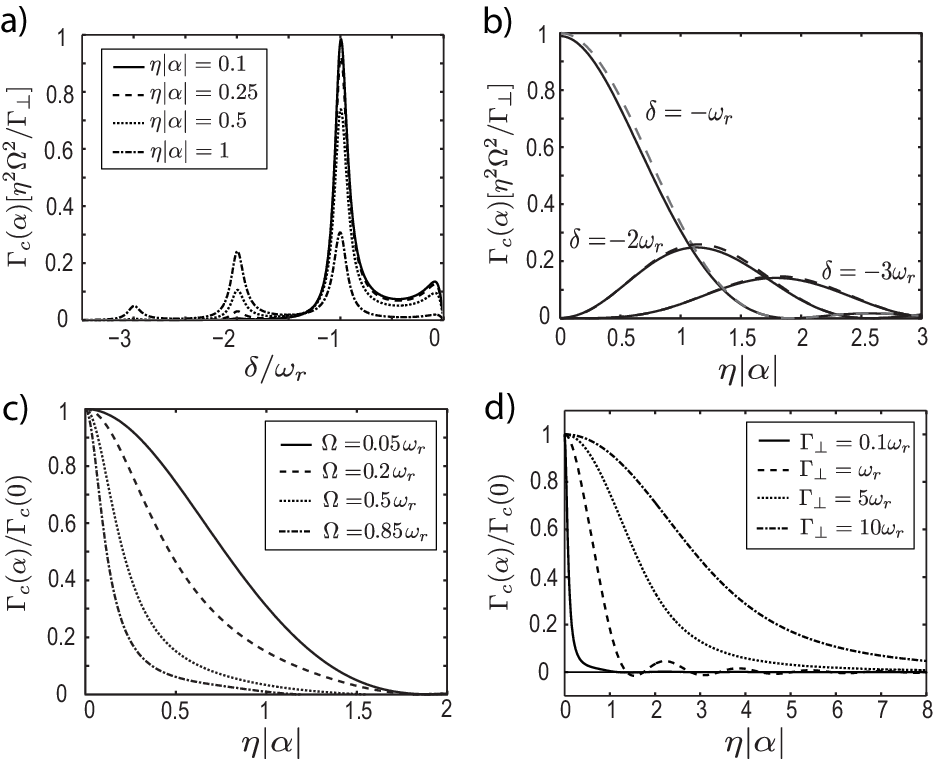}
\caption{a) Cooling rate for a weakly driven TLS for different oscillation amplitudes $\alpha$ for $\Omega=0.05\omega_r$ and  $\Gamma_\perp=0.15\omega_r$. b) For the same parameters the cooling rate is plotted for fixed detuning as a function of $\alpha$. The dashed lines indicate the analytic results given in Eq.~\eqref{eq:GammacAnalytic}. c) The cooling rate $\Gamma_c(\alpha)$ is plotted for different values of the Rabi frequency $\Omega$, $\delta=-\sqrt{\omega_r^2-\Omega^2}$ and other parameters as above. d) Dependence of $\Gamma_c(\alpha)$ on $\alpha$ for different values of $\Gamma_\perp$.  The values of $\Omega$ and $\delta$ have been chosen to optimize $\Gamma_c(0)$. 
In all plots we have assumed $\Gamma_\parallel=N_{\rm TLS}=0$.
 }\label{fig:CoolingRatesSB}
\end{center} 
\end{figure}

From Eq.~\eqref{eq:ContFrac} and Eq.~\eqref{eq:SlowCooling} we first of all see that up to second order in $\lambda$ our approach reproduces the LD cooling rate and $\Gamma_c(\alpha\rightarrow 0)\equiv\Gamma_c$ (see  Eq.~\eqref{eq:AppCoolingRateLD} in App.~\ref{app:A}). The amplitude dependence of $\Gamma_c(\alpha)$ is summarized in Fig.~\ref{fig:CoolingRatesSB} for different parameters and assuming $\Gamma_\parallel,N_{\rm TLS}\rightarrow0$. In Fig.~\ref{fig:CoolingRatesSB} a) $\Gamma_c(\alpha)$ is plotted for sideband resolved and weak driving conditions as a function of $\delta$ and $|\alpha|$.  For resonant detuning $\delta=-\omega_r$ we find a degrading of the cooling rate with increasing $|\alpha|$, which is expected from a saturation of the TLS. At larger amplitudes $\eta|\alpha| > 0.1$ we also observe a strong modification of the cooling rates as a function of $\delta$ and the appearance of additional peaks at $\delta=-2\omega_r,-3\omega_r,\dots$ which correspond to multi-phonon transitions.
For a better understanding of these features we change to the displaced oscillator basis introduced in Eq.~\eqref{eq:Hpolaron} where the driving term in the Hamiltonian takes the form 
\begin{equation}
\tilde H_{\Omega}= \frac{\Omega}{2} \left(\sigma_+ e^{\eta (a^\dag -a)} + \sigma_- e^{-\eta (a^\dag -a)}\right).
\end{equation}
By expanding the exponentials we see that this Hamiltonian couples different phonon number states $|n\rangle$ and for $n\gg1$ the $\Delta n$-phonon transition matrix element is given by~\cite{CrispPRA1992}
\begin{equation}
\begin{split}
\langle 1, n\!-\!\Delta n | \tilde H_\Omega|0, n \rangle =&\frac{\Omega}{2} e^{-\frac{\eta^2}{2}} \sqrt{\frac{(n\!-\!\Delta n)!}{n!}} \eta^{\Delta n} L_{n-\Delta n}^{\Delta n}(\eta^2)\\
\simeq  &\frac{\Omega}{2} J_{\Delta n}(2 \eta \sqrt{n}).
\end{split}
\end{equation} 
Here $L_m^k(x)$ denotes the generalized Laguerre polynomial and $J_k(x)$ is the k-th order Bessel function.
In the weak excitation limit $\Omega\rightarrow0$ the cooling rate is proportional to the square of these matrix elements and for a detuning  $\delta=-\Delta n\times \omega_r$ we obtain a cooling rate
\begin{equation}\label{eq:GammacAnalytic}
\Gamma_c(\alpha)\simeq \eta^2 \frac{\Omega^2}{\Gamma_\perp} \times\Delta n\times  \left(\frac{J_{\Delta n}(2\eta |\alpha|)}{\eta |\alpha|}\right)^2.
\end{equation}
Fig.~\ref{fig:CoolingRatesSB} b) shows that these analytic estimates correctly capture the dependence of $\Gamma_c(\alpha)$ in the weak driving limit. 

In Sec.~\ref{sec:LambDicke} we have seen that in the LD regime cooling rates  are optimized under strong driving conditions $\Omega\sim \omega_r$, and in Fig.~\ref{fig:CoolingRatesSB} c) we study $\Gamma_c(\alpha)$ for increasing Rabi frequencies. While the general dependence on  the mean occupation number is similar to the weak driving limit, the cooling rates decrease much faster with increasing $\alpha$ and the benefit of strong driving is less significant at larger oscillation amplitudes. We attribute this additional suppression of the cooling rate to a deviation from the initial resonance condition, $\delta=-\sqrt{\omega_r^2-\Omega^2}$, as the effective Rabi frequency changes with $\alpha$. This effect is less severe for larger linewidths which can be seen in Fig.~\ref{fig:CoolingRatesSB} d) where we plot $\Gamma_c(\alpha)$ for different TLS decay rates $\Gamma_\perp$ and with $\Omega$ and $\delta$ chosen to optimize $\Gamma_c(0)$.  As we go from the sideband to the Doppler regime the cooling rates become more robust and the results from the LD regime are valid up to $\lambda \alpha \lesssim \Gamma_\perp$. Fig.~\ref{fig:CoolingRatesSB} d) also shows that for $\Gamma_\perp\sim \omega_r$ and for certain values of $\alpha$ even small negative cooling rates can occur.

In summary we find a significant reduction of the cooling rate already at moderately high occupation numbers 
$\eta \sqrt{N_{th}}\sim 0.1-1$. This degrading is most sever in the sideband resolved and strong driving regime, where 
in the LD limit cooling is optimized.  However, for most parameters and as long as $\delta<0$ we obtain  $\Gamma_c(|\alpha|)>0, \forall \alpha$ and an isolated resonator mode would still be gradually cooled towards the groundstate. Therefore, to understand the meaning of a temperature dependent cooling rate in a realistic setup we must study the full dynamics of the resonator mode and include environment induced re-thermalization processes.

\section{Fokker-Planck equation}\label{sec:FokkerPlanck}
In the previous section we have discussed the amplitude dependence of the cooling rate $\Gamma_c(\alpha)$ under the assumption that the resonator mode is prepared in a coherent state $|\alpha\rangle$. We now generalize this analysis to study the evolution of arbitrary resonator states $\rho_r(t)$.  A natural way to do so is in terms of a coherent state representation for the density operator and in the present case a suitable choice is the Wigner function $W(\alpha,t)$~\cite{QRT}.  At high temperatures $W(\alpha=x+ip,t)$ corresponds to the classical probability distribution for position $x$ and momentum $p$. In a frame rotating with $\omega_r$ the Wigner function evolves according to the Fokker-Planck equation 
%\begin{equation}\label{eq:Dgamma}
%\begin{split}
%\dot W(\alpha,t)= &\frac{\gamma}{2}\left(\frac{\di}{\di \alpha} \alpha +\frac{\di}{\di \alpha^*} \alpha^*\right) W(\alpha,t)\\ +&\frac{\gamma}{2} (2N_{th}+1) \frac{\di^2}{\di \alpha \di \alpha^*} W(\alpha,t),
%\end{split}
%\end{equation}
\begin{equation}\label{eq:Dgamma}
\begin{split}
\dot W(\alpha,t)= &\frac{\gamma}{2}\left(\frac{\di}{\di \alpha} \alpha \!+\!\frac{\di}{\di \alpha^*} \alpha^*\!+\!(2N_{th}\!+\!1) \frac{\di^2}{\di \alpha \di \alpha^*}\right) W(\alpha,t)
\end{split}
\end{equation}
with a steady state $W(\alpha, \infty)=e^{-|\alpha|^2/(N_{th}+1/2)}/(\pi (N_{th}+1/2))$. To include the dynamics of the TLS  we introduce three additional distributions
\begin{equation}\label{eq:Wk}
W_{k} (\alpha,t)= \frac{1}{\pi^2} \int d^2 \beta    e^{-i\alpha \beta^*} e^{-i\alpha^* \beta}   {\rm Tr}\{ e^{i\beta a^\dag+i\beta^* a} \sigma_k \rho(t) \},
\end{equation}
where $k=-,+,z$ and  $\langle \sigma_k\rangle(t)=\int d^2\alpha \,W_k(\alpha,t)$. The $W_k(\alpha,t)$ evolve  on a timescale $T_1$ and  for $\lambda\ll \Gamma_\perp,\omega_r$ we can follow the arguments presented in Sec.~\ref{sec:HighTCoolingRates} to derive an effective equation for the much slower dynamics of $W(\alpha,t)$. The details of this derivation is outlined in App.~\ref{app:B} and as a result we obtain the modified Fokker-Planck equation 
\begin{equation}\label{eq:FPEquation}
\begin{split}
&\dot W(\alpha,t)\simeq  \frac{1}{2} \left(\frac{\di}{\di \alpha} \alpha [ i\Delta(\alpha)+\Gamma(\alpha)]  + {\rm H.c.} \right) W(\alpha,t) \\
&+\left( D(\alpha)\frac{\di^2}{\di \alpha \di \alpha^*} +\frac{M(\alpha)}{2}  \frac{\di^2}{\di \alpha^2 }  +  \frac{M^*(\alpha)}{2} \frac{\di^2}{\di \alpha^{*2} } \right)W(\alpha,t).
\end{split}
\end{equation}
%
%\begin{equation}\label{eq:FPEquation}
%\begin{split}
%&\dot W(\alpha,t)\simeq \Big[ \frac{i}{2} \left(\frac{\di}{\di \alpha} \alpha -\frac{\di}{\di \alpha^*}\alpha^*    \right) \Delta(\alpha) + \frac{1}{2}\left(\frac{\di}{\di \alpha} \alpha +\frac{\di}{\di \alpha^*}\alpha^*    \right) \Gamma(\alpha) \\
%&+ D(\alpha)\frac{\di^2}{\di \alpha \di \alpha^*} +\frac{M(\alpha)}{2}  \frac{\di^2}{\di \alpha^2 }  +  \frac{M^*(\alpha)}{2} \frac{\di^2}{\di \alpha^{*2} } \Big] W(\alpha,t).
%\end{split}
%\end{equation}
This equation describes the evolution of a damped resonator where the damping rate $\Gamma(\alpha)=\gamma+ \Gamma_c(\alpha)$,
 the frequency shift $\Delta(\alpha)$ and the diffusion terms $D(\alpha)$ and $M(\alpha)$ depend explicitly on the amplitude of the oscillation.  The rate $\Gamma_c(\alpha)$ is identical to the semiclassical cooling rate discussed above and is defined in Eq.~\eqref{eq:SlowCooling}. In the limit $\alpha\rightarrow 0$ we obtain for the diffusion terms $D(\alpha)=\gamma (N_{th}+1/2)+\Gamma_c (N_0+1/2)$ and $M(\alpha)=0$, such that Eq.~\eqref{eq:FPEquation} correctly reproduces the cooling results in the LD limit. For general $\alpha$ the amplitude dependence of $\Delta(\alpha)$, $D(\alpha)$ and $M(\alpha)$ can be calculated numerically as described in App.~\ref{app:B}.

The parameters $\Delta(\alpha)$, $\Gamma(\alpha)$ and $D(\alpha)$ depend on $r=|\alpha|$ only while $M(\alpha)\sim \alpha ^2 \tilde M(|\alpha|)$. Therefore, Eq.~\eqref{eq:FPEquation} preserves the radial symmetry of the initial thermal distribution function and we can restrict the following discussion to the radial equation,
\begin{equation}\label{eq:RadialFP}
\begin{split}
\dot W(r,t)= & \frac{1}{2}\left(\frac{\di}{\di r} r +1  \right) \Gamma(r)W(r,t)\\
 +&\frac{1}{4r}\frac{\di}{\di r} r \Gamma(r) \left(N (r)+\frac{1}{2}\right)  \frac{\di}{\di r}W(r,t).
\end{split}
\end{equation}
Here we have made use of the fact that to lowest order in $\eta$ the functions $ D(r)$ and $M(r)$ commute with the radial derivatives and we have defined 
\begin{equation}
N(r):=\frac{D(r)+{\rm Re}\{M(r)\}}{\Gamma(r)}-\frac{1}{2}.
\end{equation} 
The steady state solution of Eq.~\eqref{eq:RadialFP} is   
\begin{equation}
W(r, t\rightarrow\infty)=\mathcal{N} e^{ - R(r)},\qquad  R(r):=  \int_0^r \frac{ 2r'\, dr'}{(N(r')\!+\!1/2)},
\end{equation}
where $\mathcal{N}$ is a normalization constant and the final  occupation number $n_f$ is given by
\begin{equation}
n_f = -\frac{1}{2}+ 2\pi  \mathcal{N} \int_0^\infty dr \, r^3 \,e^{ - R(r)} .
\end{equation}

\subsection{Final occupation number} 
\begin{figure}
\begin{center}
\includegraphics[width=0.45\textwidth]{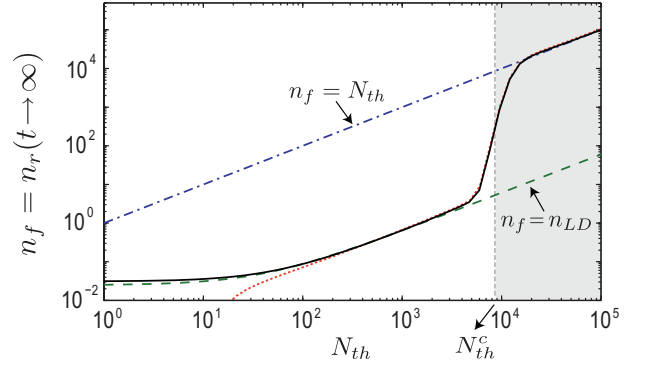}
\caption{Steady state occupation number $n_f$ as a function of the thermal equilibrium value $N_{th}$. The parameters used for this plot are $\Omega=0.5\omega_r$, $\delta=-\sqrt{\omega_r^2-\Omega^2}$, $\Gamma_\perp=0.5\omega_r$, $\eta=0.1$ and $\gamma/\omega_r=0.25\times 10^{-5}$. The value of $N_{th}^c$ indicates the crossover from the LD regime, $n_f=n_{LD}$ to a regime where cooling is strongly suppressed (shaded area) and $n_f=N_{th}$. The dotted line shows the approximate analytic expression for $n_f$  given in Eq.~\eqref{eq:HighTnf}. }\label{Figure4}
\end{center} 
\end{figure}
In Fig.~\ref{Figure4} we plot the characteristic dependence of the final resonator occupation number $n_f$ as a function of the initial value $N_{th}$. For very low $N_{th}$ the final occupation number is limited by the intrinsic limit $n_f=N_0$ and increases linearly  $n_f\approx \gamma/\Gamma_c(0)\times N_{th}$ for moderate $N_{th}$. This behavior is already captured by the LD result, $n_f\simeq n_{LD}$ as discussed in Sec.~\ref{sec:LambDicke}. However, at very high $N_{th}$ we see a crossover to a regime where cooling is no longer possible and $n_f\approx N_{th}$. 

\begin{figure}
\begin{center}
\includegraphics[width=0.45\textwidth]{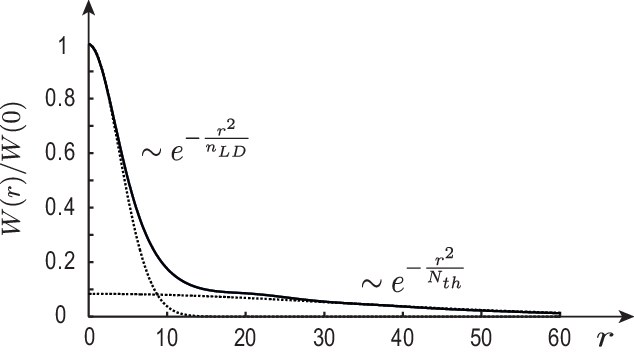}
\caption{Steady state Wigner-function showing the bimodal structure in the crossover regime $N_{th}\sim N_{th}^c$. The parameters for this plot are  $\Omega/\omega_r=0.1$, $\delta=-\omega_r$, $\Gamma_\perp=0.15\omega_r$, $\lambda=0.1$, $\gamma/\omega_r=10^{-5}$ and $N_{th}=2000$. The dotted lines indicate the LD result and a thermal equilibrium distribution. }\label{Figure5}
\end{center} 
\end{figure}

To study this crossover in more detail we consider the relevant case where diffusive terms are dominated by thermal noise such that $N(r)+1/2\approx \gamma N_{th}/\Gamma_c(r)$. Indeed we find numerically that $(D(r)+{\rm Re} \{M(r)\}) \leq D(0)$, such that this assumptions is usually well justified, in particular for large $r$.  Then
\begin{equation}
R(r)\approx \frac{r^2}{N_{th}} + \frac{2}{n_{LD}} \int_0^r dr' \, r'  \tilde \Gamma_c(r'),
\end{equation}
where $\tilde \Gamma_c(r)=\Gamma_c(r)/\Gamma_c(0)$ and $n_{LD}\simeq\gamma N_{th}/\Gamma_c(0)>1$ is the final occupation number in the LD limit. From the discussion in Sec.~\ref{sec:HighTCoolingRates} we know that $\Gamma_c(r<\eta)\simeq \Gamma_c(0)$ and in the limit $N_{th}\gg n_{LD}$ we obtain $R(r < \eta)\simeq r^2/n_{LD}$. In the opposite limit we can approximately write $R(r\gg 1/\eta )\approx r^2/N_{th} + \mathcal{I}_1/(\eta^2 n_{LD})$ where  
\begin{equation} 
\mathcal{I}_1=2 \int_0^\infty dx \, x \, \tilde \Gamma_c\left(r=\frac{x}{\eta}\right),
\end{equation}
is a numerical constant $\sim \mathcal{O}(1)$. Therefore, the steady state can be approximately written as the sum of two distributions,
\begin{equation}\label{eq:Bimodal}
W(r)\approx \mathcal{N} \left(  e^{-\frac{\mathcal{I}_1}{\eta^2 n_{LD}}} e^{ - \frac{r^2}{N_{th}}} + e^{-\frac{r^2}{n_{LD}}}\right),
\end{equation}
as illustrated in Fig.~\ref{Figure5}. For $\eta^2 n_{LD} < \mathcal{I}_1$ the steady state distribution corresponds to the LD result while 
in the opposite case $ \eta^2 n_{LD} > \mathcal{I}_1$ it approaches the thermal equilibrium. Therefore, while a degrading of cooling rates already occurs at occupation numbers $N_{th}\sim1/\eta^2$ the final state of the resonator is still described by the LD theory as long as the much weaker condition $n_{LD}<\mathcal{I}_1/\eta^2$ is satisfied. In particular, this implies that -- within the validity of our model -- ground state cooling is always correctly predicted by the LD approximation. 

From the approximate bimodal steady state distribution given in Eq.~\eqref{eq:Bimodal} we can calculate the final occupation number
\begin{equation}\label{eq:HighTnf}
n_f \approx \left( \frac{1}{\zeta_{LD}} + \frac{1-1/\zeta_{LD}}{1+e^{\frac{\mathcal{I}_1 \zeta_{LD}}{N_{th}\eta^2}}/\zeta_{LD}}\right)\times N_{th},
\end{equation}
where we have introduced the LD cooling factor $\zeta_{LD}=\Gamma_c(0)/\gamma \gg 1$. From this result we can deduce a critical thermal occupation number 
\begin{equation}\label{eq:Nc1}
N^{c}_{th}  = \frac{ \mathcal{I}_1}{ \eta^2} \times \frac{ \zeta_{LD}}{\log(\zeta_{LD})}, 
\end{equation}
beyond which cooling is suppressed. In Eq.~\eqref{eq:Nc1} $N_{th}^c$ is written in terms of $\eta$ and a given cooling factor $\zeta_{LD}$. However, both quantities scale like $\sim\lambda^2$ and alternatively we can rewrite $N_{th}^c$ as 
\begin{equation}\label{eq:Nc2}
N^{c}_{th}  = Q \times \frac{\omega_r}{\Gamma_\perp} \times \frac{\mathcal{I}_2}{\log(\zeta_{LD})},
\end{equation}
where $Q=\omega_r/\gamma$. Here we have introduced a single numerical constant $\mathcal{I}_2=(\Gamma_c(0)\Gamma_\perp/\lambda^2)\times \mathcal{I}_1$ which contains information about the normalized cooling rate at $\alpha=0$ as well as its temperature dependence. From Eq.~\eqref{eq:Nc2} we see that $N_{th}^c$ has only a weak logarithmic dependence on the coupling strength $\lambda$.

\subsection{Discussion}

For a weakly driven TLS we can use the analytic expression given in Eq.~\eqref{eq:GammacAnalytic} to show that $\mathcal{I}_1=1$ and $\mathcal{I}_2\approx \Omega^2/\omega_r^2$. For strong driving $\Omega \sim \omega_r$ and $\Gamma_\perp\sim \omega_r$ we obtain numerical values in the range of $\mathcal{I}_2\approx 0.1-0.2$ which decrease again for $\Gamma_\perp \ll \omega_r$ and $\Gamma_\perp \gg \omega_r$. Therefore, assuming $\Gamma_\perp\sim \omega_r$ Eq.~\eqref{eq:Nc2} states that cooling of the mechanical mode with a TLS is only possible if the initial equilibrium occupation number is about 10-100 times smaller than the mechanical quality factor $Q$. For  cryogenic temperatures of $T\leq 100$ mK and mechanical frequencies of $\omega_r/2\pi \geq 10$ MHz we obtain $N_{th}<500$ and for reasonable Q-values of $Q\sim10^5$ non-linear effects do not play a role.  However,  at  $T=4$ K or frequencies $\omega_r/2\pi \leq 1$ MHz, the thermal occupation number $N_{th}\approx 10^4$ can be comparable with $N_{th}^c$.  Under these condition Eq.~\eqref{eq:HighTnf} must be used to determine the final occupation number for the specific set of experimental parameters. Finally, we see that at room temperature cooling will be difficult to achieve and requires the combination of high mechanical frequencies with exceptionally high mechanical quality factors.

\section{Multi-mode effects}\label{sec:MultiMode}
Our analysis so far has been based on the model Hamiltonian~\eqref{eq:H} where the mechanical resonator is approximated by a single vibrational mode, e.g., the fundamental bending mode. In reality the TSL is also coupled to higher order vibrational modes and more accurately  
\begin{equation}\label{eq:HMM}
H= -\frac{\delta}{2}\sigma_z + \frac{\Omega}{2}\sigma_x + \sum_{k}\omega^k_r a_k^\dag a_k +\sum_k \frac{\lambda_k}{2} (a_k+a_k^\dag)\sigma_z\,.
\end{equation}
Here $k=0$ is the fundamental mode, $a_0\equiv a$, which we have studied so far. The $a_{k>0}$ are operators for higher order mechanical modes of  frequency $\omega_r^k$ which are coupled to the TLS with a strength $\lambda_k$. To provide a rough estimate of the influence of these higher order modes on the cooling rate of the fundamental mode let us switch to the displaced oscillator basis introduced in Eq.~\eqref{eq:Hpolaron}, but now with respect to all modes. We obtain
\begin{equation}\label{eq:HpolaronMulti}
\tilde H_{\Omega}=\frac{\Omega}{2}\left(e^{\sum_k \eta_k (a_k^\dag-a_k)}\sigma_+ +e^{-\sum_k \eta_k(a_k^\dag-a_k)} \sigma_-\right),
\end{equation}
where $\eta_k=\lambda_k/\omega_r^k$ is the LD for the $k$-th mode. For system parameters which are optimized to cool the fundamental mode we can assume that non of the higher order modes is resonantly driven and therefore these modes approximately remain in a thermal state. Then, by averaging Eq.~\eqref{eq:HpolaronMulti} over the thermal distribution of $k>0$ modes we obtain a reduced Rabi-frequency $\Omega\rightarrow \Omega \langle e^{\sum_{k>0} \eta_k (a_k^\dag-a_k)}\rangle_0$ and a corresponding reduction of the zeroth mode cooling rate 
\begin{equation}\label{eq:ReducedRc}
 \Gamma_c(\alpha)\rightarrow \Gamma_c(\alpha)  e^{-\sum_{k>0} \eta_k^2(2N_{th}^k+1)}= \Gamma_c(\alpha) e^{- \beta \eta_0^2 N^0_{th}}.
\end{equation}
Here we have expressed the reduction factor in terms of the LD parameter and equilibrium occupation number of the fundamental mode and a numerical constant $\beta=2\sum_{k>0} (\lambda_k/\lambda_0)^2\times(\omega_r^0/\omega_r^k)^3$. It depends on the geometry of the mechanical beam~\cite{ElasticityBook} and the exact coupling mechanism. For a TLS with a point-like coupling to the tip of a cantilever we obtain $\beta\approx 0.0085$, and $\beta\approx 0.011$ for a TLS coupled to the center of a doubly clamped beam with low tensile stress. In both cases $\omega_k\sim k^2$ and only one or two high frequency modes contribute to the sum.  For a high stress doubly clamped beam we obtain a slightly higher value of $\beta\approx 0.096$, since the mode frequencies increase only linearly with $k$.

The approximated cooling rate given in Eq.~\eqref{eq:ReducedRc} shows that a single mode model for a mechanical beam coupled to a TLS is only valid for $N_{th}<1/(\beta \eta^2)$. For higher temperatures  the thermal motion of higher order vibrational modes has a significant influence on the system dynamics and in principle a full multi-mode model must be considered. However, the influence of high frequency mode decreases quickly with increasing $\omega_r^k$ which suggest that, e.g., a two mode cooling scheme or an elaborate coupling design can be sufficient to eliminate these effects.

\section{Summary $\&$ conclusions}\label{sec:Conclusions}

In this work we have analyzed the role of non-linear effects on the cooling dynamics of a mechanical resonator  coupled to a dissipative TLS. We have found that already at moderately high occupation numbers $N_{th}\sim \eta^{-2}$ a significant reduction of the cooling rate can occur. The reduction is most pronounced in the strong-driving and sideband resolved regime, where in the low temperature limit cooling is optimized. For the steady state we have found a crossover from the LD to the high temperature regime where cooling of a resonator mode with a single TLS is no longer possible.  We have derived an approximate expression for the final occupation number which accurately describes this crossover and discussed the dependence of the critical occupation number $N_{th}^c$ on the relevant system parameters. Our results are relevant for experimental implementations of cooling schemes operating at temperatures above $\sim1 $K and in general for cooling of resonator modes with frequencies of $\sim1$ MHz and below.

The theoretical approach presented in this paper provides a unified description of the dynamics of a weakly coupled resonator-TLS system in the quantum as well as the high temperature regime.  It is valid for $\lambda \ll \Gamma_\perp, \omega_r$ which is fulfilled for many solid state TLS coupled to nano-mechanical resonators. Therefore, apart from cooling our analysis can be adopted to study related problems like non-linear dissipation mechanisms caused by single or few two-level fluctuators~\cite{RemusPRB2009} or phonon lasing physics~\cite{HaussPRL2008}.

\subsection*{Acknowledgments}
The author thanks I. Wilson-Rae, S. Kolkowitz, A. Jayich, M. Lukin and J. Harris for stimulating discussions and valuable feedback on this project. This work is support by the NSF through a grant for the Institute for Theoretical Atomic, Molecular and Optical Physics at Harvard University and Smithsonian Astrophysical Observatory.

%%%%%%%%%%%%%%%%%%%%%%%%%%%%%%%%%%%%%%%%%%%%%%%%%%
%   
%                                                                  Appendix
%
%%%%%%%%%%%%%%%%%%%%%%%%%%%%%%%%%%%%%%%%%%%%%%%%%%

\begin{appendix}

\section{The fluctuation spectrum $S(\omega)$}\label{app:A} 
To evaluate the fluctuation spectrum $S(\omega)$ defined in Eq.~\eqref{eq:Spectrum} we write $ S(\omega)= \lambda^2/2\times  {\rm Re}\{ \vec C_3(s=-i\omega)\}$ where the vector $\vec C(s)$ is the Laplace transform of the set of correlation functions 
$
\vec C(\tau)= \langle \vec S(\tau) \sigma_z\rangle_0- \langle \vec S\rangle_0\langle \sigma_z\rangle_0
$
and $\vec S=(\sigma_-, \sigma_+, \sigma_z)^T$. Using the quantum regression theorem~\cite{QRT} we obtain
\begin{equation}
\vec C(s) =  \frac{1}{ s \mathbbm{1}-  {\bf A}}  \left[\left(\begin{array}{c}
\langle \sigma_-\rangle_0    \\
-\langle \sigma_+\rangle_0   \\
1 
\end{array}\right) - \langle \vec S  \rangle_0 \langle \sigma_z\rangle_0  \right],
\end{equation}
where the matrix ${\bf A}$ is defined in Eq.~\eqref{eq:AMatrix} and the steady state expectation values $\langle \vec S\rangle_0$ follow from the Bloch equations of the TLS,  $\langle \dot {\vec S}\rangle = {\bf A}\langle  \vec S\rangle - \Gamma_\perp \vec V_z$, where $ \vec V_z=(0,0,1)^T$.
The cooling rate can be further simplified and written as
\begin{equation}\label{eq:AppCoolingRateLD}
\Gamma_c= -\lambda^2 {\rm Re} \, \Big\{ (0,0,1) \frac{1}{i\omega_r\mathbbm{1}+{\bf A}} \left(\begin{array}{c}
\langle \sigma_-\rangle_0    \\
-\langle \sigma_+\rangle_0   \\
0 
\end{array}\right) \Big\},
\end{equation}
which agrees with the result derived in Sec.~\ref{sec:HighTCoolingRates} for $\Gamma_c(\alpha\rightarrow 0)$.

\section{Fokker-Planck equation}\label{app:B}
In this appendix we derive the effective Fokker-Planck equation~\eqref{eq:FPEquation}. We change into a frame rotating with the mechanical frequency $\omega_r$ such that the free evolution of the resonator Wigner-function $W(\alpha,t)$ is $\dot W(\alpha,t)=\mathcal{D}_\gamma W(\alpha,t)$. The differential operator for mechanical damping $\mathcal{D}_\gamma$ is defined by Eq.~\eqref{eq:Dgamma}. The distributions $W_k(\alpha,t)$ introduced in Eq.~\eqref{eq:Wk} are grouped into a vector $\vec{\mathcal{W}}= (W_-,W_+,W_z)^T$ which for $\lambda\rightarrow 0$ evolves as
\begin{equation}
\dot{\vec{\mathcal{W}}}(\alpha,t)= {\bf A}\vec{\mathcal{W}}(\alpha,t) -\Gamma_\perp \vec V_z W(\alpha,t) +\mathcal{D}_\gamma \vec{\mathcal{W}}(\alpha,t).
\end{equation}
For finite $\lambda$ the evolution of $W(\alpha)$ and $\vec{\mathcal{W}}(\alpha)$ is coupled by the TLS-resonator interaction 
$
H_\lambda(t)=\frac{\lambda}{2}(a e^{-i\omega_rt} +a^\dag e^{i\omega_r t}) \sigma_z,
$
and we obtain the additional terms  
\begin{equation}
\begin{split}
\dot W_{\sigma_k}(\alpha)= &- i \frac{\lambda}{2}\left(\alpha  e^{-i\omega_r t} +\alpha^*  e^{i\omega_r t}\right)W_{[\sigma_k,\sigma_z]}(\alpha) \\
&- i \frac{\lambda}{4}\left( e^{-i\omega_r t} \frac{\di}{\di \alpha^*} - e^{i\omega_r t} \frac{\di}{\di \alpha}\right) W_{\{\sigma_k,  \sigma_z\}_+}(\alpha),
\end{split}
\end{equation}
where $\{.,.\}_+$ is the anti-commutator.  To remove the explicit time dependence we change to Floquet representation,
\begin{equation}
W_k(\alpha,t)= \sum_{n=-\infty}^\infty W_k^n(\alpha,t) e^{-i\omega_r n t},
\end{equation}
and obtain the set of coupled equations 
\begin{equation}\label{eq:WnRes}
\begin{split}
&\dot W^n(\alpha,t) = i\omega_r n W^n(\alpha,t)+\mathcal{D}_\gamma W^n(\alpha,t)\\&+  i \frac{\lambda}{2} \left( \frac{\di}{\di \alpha} W^{n+1}_{z}(\alpha,t) - \frac{\di}{\di \alpha^*} W^{n-1}_{z}(\alpha,t)\right),
\end{split}
\end{equation}
and
\begin{equation}\label{eq:B5}
\begin{split}
\dot{\vec{\mathcal{W}}}^n(\alpha,t)&= {\bf A}_n \vec{\mathcal{W}}^n(\alpha,t) -\Gamma_\perp \vec V_z W^n(\alpha,t) +\mathcal{D}_\gamma  \vec{\mathcal{W}}^n(\alpha,t)\\&- i \lambda \alpha {\bf A}_z  \vec{\mathcal{W}}^{n-1}(\alpha,t) - i \lambda  \alpha^*  {\bf A}_z \vec{\mathcal{W}}^{n+1}(\alpha,t)\\
&- i \frac{\lambda}{2} \vec V_z  \left(  \frac{\di}{\di \alpha^*}  W^{n-1}(\alpha,t) -\frac{\di}{\di \alpha} W^{n+1}(\alpha,t)\right),
%&+ i \frac{\lambda}{2}  \langle \sigma_z\rangle_0 \left(  \frac{\di}{\di \alpha^*}   \vec{\mathcal{W}}^{n-1}(\alpha,t) -\frac{\di}{\di \alpha}  \vec{\mathcal{W}}^{n+1}(\alpha,t)\right),
\end{split}
\end{equation}
where we have defined  ${\bf A}_n:={\bf A}+i\omega_r n \mathbbm{1}$. Our goal is to eliminate the fast dynamics of the TLS and to
derive an effective Fokker-Planck equation for the slowly varying Wigner-function $W^0(\alpha,t)$. To do so we point out that 
\begin{equation}
\lambda \frac{\di}{\di \alpha} W_k(\alpha,t)\sim \frac{ \lambda\alpha}{n_r(t)}  W_k(\alpha,t)\sim \mathcal{O}(\lambda) \times  W_k(\alpha,t),
\end{equation}
where $n_r(t)$ is the mean resonator occupation number. Therefore, in the regime $\lambda \ll \Gamma_\perp,\omega_r$ we can formally expand the coupled differential equations in powers of $\lambda \times  \di/\di \alpha$.

\emph{Cooling rates.} To zeroth order in $\lambda \times  \di/\di \alpha$ we obtain for the resonator mode $\dot W^0(\alpha,t)=\mathcal{D}_\gamma W^0(\alpha,t)$ and $W^n(\alpha,t)=0$ otherwise.  In Eq.~\eqref{eq:B5} we must include terms $\sim \alpha \lambda$ and to zeroth order in our expansion parameter we obtain $W^n_k(\alpha,t)=S^n_{k,0}(\alpha)\times W^0(\alpha,t)$. Here the $S^n_{k,0}(\alpha)$ have been introduced in Sec.~\ref{sec:HighTCoolingRates} and they are defined by the matrix continued fraction in Eq.~\eqref{eq:ContFrac}. Note that in deriving this expression we have neglected the influence of mechanical damping. If the resonator state is close to thermal equilibrium $\mathcal{D}_\gamma W_k \sim \mathcal{O}(\gamma) $, while far away from equilibrium $\mathcal{D}_\gamma W_k \sim \mathcal{O}(N_{th}\gamma) $. However, to obtain a non-equilibirum state we require $\gamma N_{th} \ll \Gamma_c < \Gamma_\perp$ and therefore as long as  $\Gamma_\perp \gg \gamma$ this approximation is automatically justified.

After inserting the zeroth order results for $W^n_k(\alpha)$ back into Eq.~\eqref{eq:WnRes} we obtain to  first order in $\lambda \times  \di/\di \alpha$, 
\begin{equation}\label{eq:W0firstorder}
\begin{split}
&\dot W^0(\alpha,t)=\mathcal{D}_\gamma W^0(\alpha,t)\\
&+ i \frac{\lambda}{2}\left[ \frac{\di}{\di \alpha} S^{+1}_{z,0}(\alpha) -\frac{\di}{\di \alpha^*} S^{-1}_{z,0}(\alpha)\right] W^0(\alpha,t).
\end{split}
\end{equation}
From this expression we can deduce a cooling rate $\Gamma_c(\alpha)$ and a frequency shift $\Delta(\alpha)$ by setting
\begin{equation}
i\lambda S^{+1}_{z,0}(\alpha) =: \alpha \left( \Gamma_c(\alpha) +i \Delta(\alpha)\right),
\end{equation}
which reproduces the semiclassical results derived in Sec.~\ref{sec:HighTCoolingRates}. 

\emph{Diffusion.} We now include corrections to $W_k^n(\alpha)$ which are first order in $\lambda \times  \di/\di \alpha$ and lead to diffusion terms in the equation for $W^0$. This requires some care and we first point out that apart from damping the zeroth order results also leads to a static force $\sim S_{z,0}^0(\alpha)$. In the LD regime $S_{z,0}^0(\alpha)=\langle \sigma_z\rangle_0$ and in Sec.~\ref{sec:LambDicke} we have eliminated this force by redefining the resonator equilibrium position. To obtain physically meaningful result the same must be done here 
and therefore we perform a variable transformation $\alpha \rightarrow \alpha +  \lambda/(2\omega_r) S_{z,0}^{0}(\alpha) e^{i\omega_r t}$. To lowest order in $\lambda$ this transformation does not change the result given in Eq.~\eqref{eq:W0firstorder}.  However, in Eq.~\eqref{eq:B5} this transformation introduces two additional terms
\begin{equation}
\begin{split}
\dot{\vec{\mathcal{W}}}(\alpha,t) = \dots  &+ i \frac{\lambda}{2}  S_{z,0}^{0}(\alpha)  \frac{\di}{\di \alpha^*}   \vec{\mathcal{W}}^{n-1}(\alpha,t) \\
&-i \frac{\lambda}{2}  S_{z,0}^{0}(\alpha) \frac{\di}{\di \alpha}  \vec{\mathcal{W}}^{n+1}(\alpha,t),
\end{split}
\end{equation}
which must be taken into account to correctly reproduce the diffusion terms found in the LD regime.

We can now insert the zeroth order result $W^n_k(\alpha,t)=S^n_{k,0}(\alpha)\times W^0(\alpha,t)$ back into the modified Eq.~\eqref{eq:B5} and iteratively solve for corrections which are linear in $\lambda \times  \di/\di \alpha$. Note that $\frac{\di}{\di \alpha} \vec S^n(\alpha)= \vec S^n(\alpha) \frac{\di}{\di \alpha}+ \mathcal{O}(\lambda)$ and to lowest order in $\lambda$ we can exchange $S_k^n(\alpha)$ and the derivatives.  Then, the final result can be written in the form  
\begin{equation}
\begin{split}
&i\lambda W_z^{+1}(\alpha,t)=  \alpha \left( \Gamma_c(\alpha) +i \Delta(\alpha)\right) W^0(\alpha,t)\\
         &+   f_2(\alpha) \frac{\di}{\di \alpha^*}  W^0(\alpha,t) +  f_3 (\alpha) \frac{\di}{\di \alpha} W^0(\alpha,t).
\end{split}
\end{equation}
After inserting this expression back into Eq.~\eqref{eq:WnRes} we end up with the Fokker-Planck equation~\eqref{eq:FPEquation}, where $D(\alpha):={\rm  Re} \{ f_2(\alpha)\}+ \gamma(N_{th}+1/2)$ and $M(\alpha):=f_3(\alpha)$. In the LD regime, i.e. up to second order in  $\lambda$,  we obtain
\begin{equation}
\begin{split}
i\lambda \vec W_z^{+1}(\alpha,t)=  -\lambda^2 \alpha (0,0,1) ({\bf A}_{+1})^{-1} {\bf A}_z \vec S_0 W^0(\alpha,t)\\
\frac{\lambda^2}{2}  (0,0,1) ({\bf A}_{+1})^{-1} \left( \langle \sigma_z\rangle_0 \vec S_0- \vec V_z \right) \frac{\di}{\di \alpha^*}W^0(\alpha,t) .
\end{split}
\end{equation}
By comparing this expression with the result derived in App.~\ref{app:A} we see that $\Gamma_c(\alpha)=\Gamma_c$ and $D(\alpha)=\Gamma_c (N_0+1/2)+ \gamma(N_{th}+1/2)$ agree with the LD results. For arbitrary $\alpha$ the coupled set of Eqs.~\eqref{eq:B5} can be evaluated numerically by truncating the coupled set of equations at a large value of $|n|$.

\end{appendix}


\begin{thebibliography}{99}
\bibitem{NoteNTLS} A finite $N_{\rm TLS}$ can be simply included by replacing $\Gamma_\perp$ by $T_1^{-1}$ and by dividing the resulting cooling rate by an additional factor $(2 N_{\rm TLS}+1)$.
\end{thebibliography}
\end{document}